\documentclass[%
aip, pof, amsmath,amssymb,
superscriptaddress,
reprint,%
onecolumn,
]{revtex4-1}

 \usepackage{tikz}
 \usetikzlibrary{positioning}
 \usetikzlibrary{backgrounds}
 \usetikzlibrary{arrows,shapes}
 \usetikzlibrary{decorations.markings}
 \usetikzlibrary{decorations.pathmorphing}
 \usetikzlibrary{decorations.pathreplacing}
 \usetikzlibrary{patterns}
 \usetikzlibrary{shapes.misc}
 \usepackage{pgfplots}
 \usepackage{pgfplotstable}
 \usepackage{multirow}

 \pgfplotsset{%
  compat=newest,
  ylabel style={rotate=-90, font=\normalsize},
  xlabel style={font=\normalsize},
  minor tick num=3,
  enlargelimits=false,
 xtick pos = left,
 ytick pos = left,
  every outer x axis line/.style={line width=0.8pt},
  every outer y axis line/.style={line width=0.8pt},
  tick style={line width=0.5pt, gray!80},
  tick align=outside,
  major tick length=0.2cm,
  minor tick length=0.1cm,
  ampmini/.style={xlabel=$1/\epsilon$, 
 		ylabel=$\log(\text{amplitude})$, 
 		ylabel style={rotate=90, font=\normalsize},
 		xlabel style={font=\normalsize},
 		width=\textwidth}
 }
 \tikzset{%
   >=stealth,
  every mark/.style={scale=0.5, color=black}}

\tikzstyle{menode}=[scale=0.7, minimum width=1.6cm, rectangle, draw, very thin, fill=gray!5, rounded corners=3pt, font=\footnotesize] 

\usepackage{microtype}
\usepackage{graphicx}
\usepackage{rotating}
\usepackage{dcolumn}
\usepackage{bm}
\usepackage{color}
\usepackage{amsmath}
\usepackage{amssymb}
\usepackage{upgreek}
\usepackage{natbib}
\usepackage[tightpage]{preview}

\begin{document}

\preprint{PoF draft version}

\title[Curvature suppresses the Rayleigh-Taylor instability]{Curvature suppresses the Rayleigh-Taylor instability}

\author{Philippe H. Trinh}%
\thanks{Corresponding author: trinh@maths.ox.ac.uk}
\affiliation{Oxford Centre for Industrial and Applied Mathematics, Mathematical Institute,\\ University of Oxford, Oxford OX2 6GG, UK}
\author{Hyoungsoo Kim}
\thanks{P. Trinh and H. Kim contributed equally to this work.}
\affiliation{Department of Mechanical and Aerospace Engineering, Princeton University,\\ Princeton, NJ 08544 USA}
\author{Naima Hammoud}
\affiliation{Department of Mechanical and Aerospace Engineering, Princeton University,\\ Princeton, NJ 08544 USA}
\author{\\Peter D. Howell}%
\affiliation{Oxford Centre for Industrial and Applied Mathematics, Mathematical Institute,\\ University of Oxford, Oxford OX2 6GG, UK}
\author{S. Jonathan Chapman}%
\affiliation{Oxford Centre for Industrial and Applied Mathematics, Mathematical Institute,\\ University of Oxford, Oxford OX2 6GG, UK}
\author{Howard A. Stone}%
\affiliation{Department of Mechanical and Aerospace Engineering, Princeton University,\\ Princeton, NJ 08544 USA}
\date{\today}

\begin{abstract}
\noindent The dynamics of a thin liquid film on the underside of a curved cylindrical substrate is studied. The evolution of the liquid layer is investigated as the film thickness and the radius of curvature of the substrate are varied. A dimensionless parameter (a modified Bond number) that incorporates both geometric parameters, gravity, and surface tension is identified, and allows the observations to be classified according to three different flow regimes: stable films, films with transient growth of perturbations followed by decay, and unstable films. Experiments and theory confirm that, below a critical value of the Bond number, curvature of the substrate suppresses the Rayleigh-Taylor instability.
\end{abstract}

\maketitle


\noindent When a dense fluid is accelerated into a fluid of lower density, the interface between the two fluids becomes unstable. For over a century, the gravitational version of this problem -- the Rayleigh-Taylor instability -- has been studied extensively, both experimentally 
and theoretically~\cite{taylor1950instability, Rayleigh1883, chandrasekhar1961hydrodynamic, drazin2004hydrodynamic}, because the phenomenon is ubiquitous in nature and technology~\cite{sharp1984overview, lindl1975two, houseman1997gravitational}. In general, these studies of gravitationally driven instabilities are performed in a planar geometry that is either horizontal or  inclined~\cite{fermigier1992two,king2007liquid, indeikina1997drop}. Here we study the Rayleigh-Taylor instability in a curved geometry and identify conditions where the curvature of the substrate suppresses the instability.

One application for film flow on a curved surface concerns the development of plasma-facing components for fusion energy reactors. In such systems it is necessary to protect the plasma and the container wall from each other. It has been suggested that a film of liquid metal offers many advantages over solids as a plasma-facing material~\cite{Majeski,Kaita,MajeskiConfProc,hazeltine2009research, shimomura2007iter}. For such a system to be successful, it must operate in a regime where the film flowing along the underside of the curved surface remains stable. 

Although most studies of thin film dynamics have considered flow along a planar substrate, there is a small literature where the ideas are developed for curved substrates. Such studies include the temporal evolution of two-dimensional thin liquid films, which exhibit thinning of the interface near regions of large curvature \cite{SchwartzWeidner}, the flow along the interior surface of a cylindrical tube with a weakly curved centerline \cite{jensen}, and the flow inside a rotating horizontal cylinder, e.g. \cite{Moffatt,Ashmore,benilov2005}. A general set of equations for viscous flow of a thin film along a curved substrate was proposed by Roy et al. \cite{RoyRobertsSimpson} and Howell \cite{Howell}, where the latter identified three distinguished limits, depending on whether an appropriate dimensionless curvature of the substrate is small, nearly constant, or large (see also \cite{MyersCharpinChapman}). A simplification of these equations is the starting point for the theory presented below.

\begin{figure}[htb]
  \centering
 \includegraphics[width=0.9\textwidth]{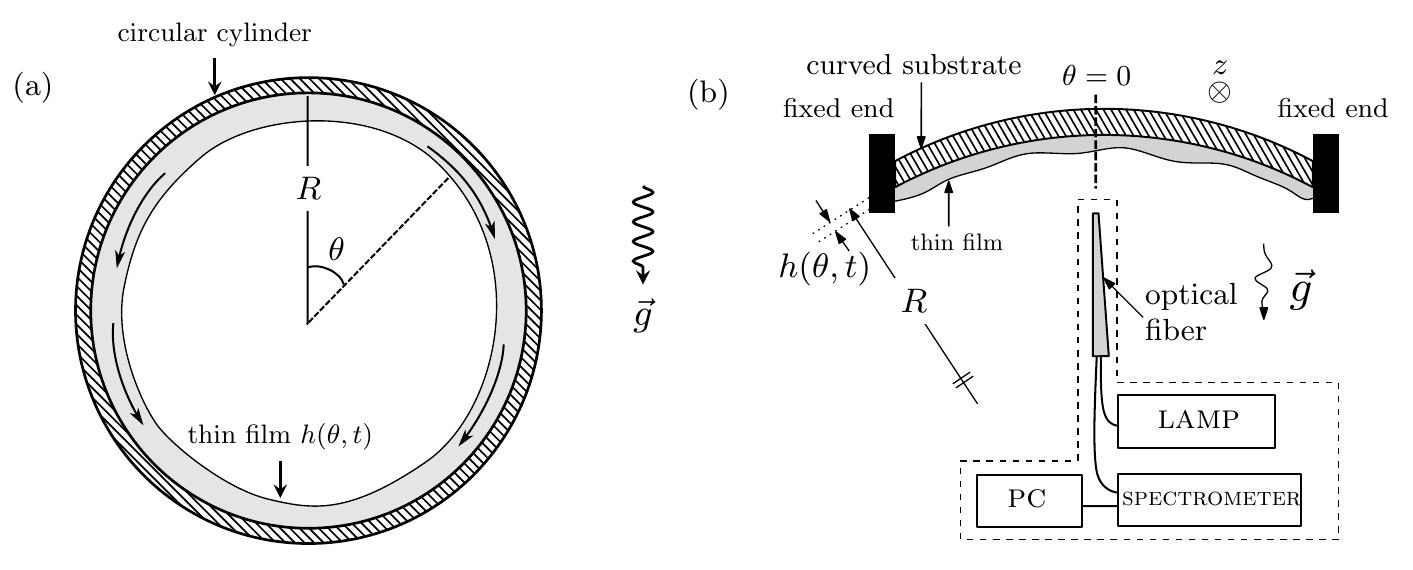}
  \caption{(a) Thin film flow on a circular cylinder and (b) schematic of the experimental setup. The substrate is formed from a curve sheet with the axial direction ($z$) directed into the page. \label{Fig:setup}}
\end{figure}

In this Letter, we study experimentally and theoretically the hydrodynamic stability of a suspended thin liquid film under a curved surface. We use the thin-film equations, well-known from viscous flow theory, to provide a description of the role of substrate curvature on the gravitationally driven flow of the film on the underside of a circular cylinder (Fig.~\ref{Fig:setup}a). These equations are solved analytically in the crucial regime near the top of the cylinder. The analytical predictions are confirmed by numerical solutions of the full equations. In addition, we perform experiments to measure the evolution of the thin film shape for different aspect ratios between the film thickness and the radius of curvature of the substrate formed from a curved sheet (Fig.~\ref{Fig:setup}b). Based upon the equations, we identify a single dimensionless parameter $B= \rho g R h_i/\gamma$, a modified Bond number, that governs the behaviour of the flow and incorporates the surface tension $\gamma$, initial film 
thickness $h_i$, and substrate curvature $R^{-1}$, where $\rho$ is the density of the fluid and $g$ is the 
gravitational acceleration. Below a critical value of $B$, the film is stable to perturbations with a thickness that decreases monotonically in time. For larger values of $B$, i.e. thicker films (or flatter substrates), the dynamics exhibit either a regime of perturbation growth followed by decay, or instability via drop formation. The combination of theory and experiments provides a relatively complete characterization of the Rayleigh-Taylor instability of a liquid layer under a curved surface, and the manner in which substrate curvature can stabilize the film.

{\it Theory.} -- We consider the flow of a thin liquid film initially placed on the underside of a curved substrate. Here, we focus on the two-dimensional case where the substrate is cylindrical with constant curvature, $1/R > 0$ (Fig.~\ref{Fig:setup}, left). However, the following theory is local in nature, and so is applicable to any 2D substrate geometry with a local maximum and non-singular curvature. The film thickness is denoted $h(\theta, t)$, with $\theta=0$ at the maximum point of the substrate. For the limit $h/R \ll 1$, with subscripts denoting derivatives, the thin film equation on the underside of a cylindrical surface is (e.g. \cite{OckendonOckendon,king2007liquid})
\begin{equation} \label{ThinFilmCurvedSubstrate1}
h_t + \frac{1}{3\mu R}\left[h^3\left (\frac{\gamma}{R}\kappa_\theta +  \rho g \left (\frac{h_\theta \cos \theta}{R} +\sin\theta\right )\right ) \right ]_\theta = 0,
\end{equation}
where $\mu$ is the liquid viscosity and $\kappa= R^{-2}\left (h_{\theta\theta}+h\right )$ is the leading-order mean curvature of the film, which accounts for the profile of the interface and the curvature of the substrate.

It is convenient to scale the film height by an initial average film thickness $h_i$ and time with a gravitational relaxation time $\mu R/\left (\rho g h_i^2\right )$. Then, Eq. (\ref {ThinFilmCurvedSubstrate1}) can be recast in dimensionless form as (we retain the same variables)
\begin{equation} \label{ThinFilmCurvedSubstrate2}
h_t + \frac{1}{3}\left[h^3\left (\frac{\delta^2}{B}\left (h_{\theta\theta\theta}+h_\theta\right ) + \delta h_\theta \cos \theta+\sin\theta\right ) \right]_\theta = 0,
\end{equation}
where $\delta = h_i/R$ is the aspect ratio of the film and $B= \rho g R h_i/\gamma$ is the modified Bond number. For many applications we expect $\delta\ll 1$.
\begin{figure}[htb]
  \centering
  \includegraphics[width=1.0\textwidth]{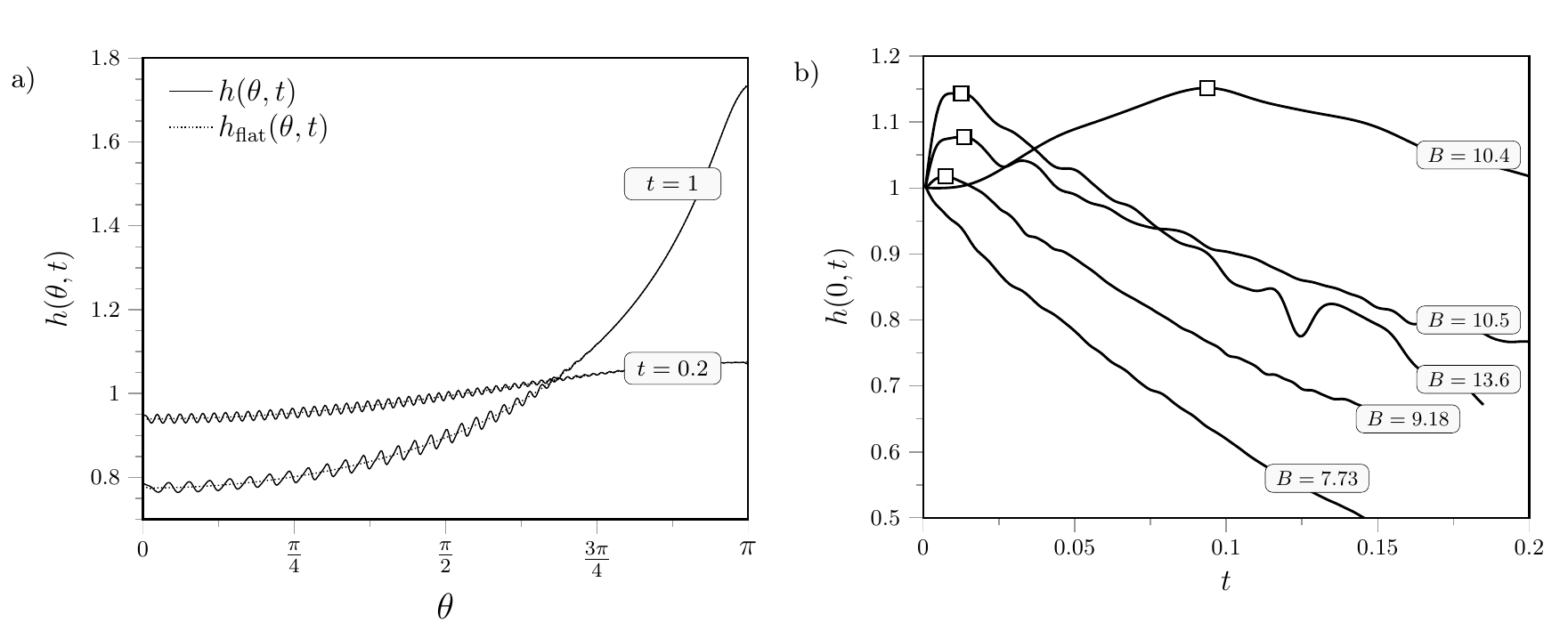}
  \caption{(a) Numerical solutions, $h(\theta, t)$ (solid), of Eq. \eqref{ThinFilmCurvedSubstrate2} at dimensionless times $t = 0.2$ and $t = 1$ for $B = 20$, $\delta = 10^{-3}$, and initial condition $h(\theta, 0) = 1 + D\sin(\pi k^* \theta/\sqrt{\delta})$, where $k^*$ is chosen to produce maximal perturbation growth and $D = 10^{-2}$ is a perturbation amplitude. The dotted lines correspond to $h_\text{flat}(\theta, t)$, the solution with $D = 0$. (b) Measurement of film thickness (re-scaled by $h_i$) as a function of non-dimensional time (re-scaled by $\mu R/(\rho g h_i^2)$) for various values of $B$. For $B > 8$, the maximal thickness is marked by a square. \label{Fig:experiments}}
\end{figure}
We are now interested in examining the growth or decay of a perturbation of the initial uniform profile, $h(\theta,0) = 1$. An example of such a profile, numerically computed using Eq. \eqref{ThinFilmCurvedSubstrate2}, is shown in Fig.~\ref{Fig:experiments}a (the details are presented in the numerical discussion to follow). We shall focus on the region near the top of the cylinder $(\theta=0)$, where the gravitational instabilities are expected to be strongest. We derive an equation that describes the local behaviour of the film by setting $\theta=\delta^{1/2}x$. Keeping only the leading-order terms in Eq. (\ref{ThinFilmCurvedSubstrate2}) for $\delta \ll 1$, we obtain a simplified equation
\begin{equation}
h_t + \frac{1}{3} \left[ h^3 \left (B^{-1} h_{xxx} + h_x + x\right )\right]_x=0.
\label{ThinFilmCurvedSubstrate3}
\end{equation}
Thus, where the destabilizing effect of gravity is expected to be the largest, only one dimensionless parameter, $B$, is significant. In the limit of zero surface tension, $B \to \infty$, Eq. (\ref{ThinFilmCurvedSubstrate3}) has the form of a backward diffusion equation, which is unconditionally unstable. This observation suggests that we should expect $B=O(1)$ for there to be any chance of stability owing to the curvature of the substrate.

With the initial condition $h(x, 0)=1$, then for any $B$, Eq. (\ref{ThinFilmCurvedSubstrate3}) has a solution $h_0(T) = T^{-1/2}$ where $T = 1+\frac{2}{3}t$. This solution corresponds to gravitationally driven drainage with uniform thinning of the film. It is natural to consider the evolution of small perturbations to this uniform profile by setting $h(x,t) = h_0 (T)\left ( 1+\epsilon \eta(x,T)\right )$, where $\epsilon \ll 1$. From Eq. (\ref{ThinFilmCurvedSubstrate3}) and at leading order in $\epsilon$, the perturbed shape, $\eta(x,T)$, satisfies 
\begin{equation} \label{ThinFilmPerturbedPDElinearized1}
\eta_T +\frac{1}{2T^{3/2}}\left (B^{-1}\eta_{xxxx} + \eta_{xx}\right ) +\frac{3}{2T}\left (x\eta_x + \tfrac{2}{3}\eta\right )=0.
\end{equation}
We consider an initial sinusoidal perturbation $\eta(x,1)=e^{ikx}$ and solve Eq. (\ref{ThinFilmPerturbedPDElinearized1}) to find
\begin{gather} \label{etaeq}
\eta(x,T)=A\left (k, T\right )\exp\left(\frac{ikx}{T^{3/2}}\right), 
\\
A(k, T) = \frac{1}{T} \exp\left[\left(1-T^{-7/2}\right)\frac{k^2}{7} - B^{-1}\left(1-T^{-13/2}\right)\frac{k^4}{13}\right].\label{Aeq}
\end{gather}
For a given initial wave number, $k$, the amplitude factor $A(k, T)$ dictates the growth or decay of a perturbation as a function of $B$ and time. A detailed analysis of this function establishes that $\eta$ will either decay monotonically, or initially grow and decay afterwards. The initial behavior of the disturbance is given by
\begin{equation} 
A\left (k,T\right )\sim 1 + \frac{\Delta }{2}\left (T-1\right ) \quad\hbox{as}~T\rightarrow 1^+,
\end{equation}
where $\Delta = k^2-2-B^{-1}k^4$. It follows that initial growth of perturbations occurs for $B > 8$, and otherwise the perturbations all decay monotonically in time. Thus, we predict that there is a critical value of the Bond number, which relates the initial film height $h_i$ and the substrate curvature $1/R$, such that the Rayleigh-Taylor instability is suppressed. According to Eq. (\ref{Aeq}), $A(k, T) \sim \text{const.}/T$ as $T \to \infty$, so eventually all (small) disturbances decay algebraically regardless of the initial wavenumber $k$. 

In summary, the conclusions of the asymptotic analysis
are: (i) any initial disturbance must immediately decay
if $B < 8$  and (ii) for $B > 8$, although any disturbance must eventually decay according to the linearized theory, the initial transient growth may result in an amplification that is exponentially large; this latter feature we expect to dominate the dynamics for large $B$. Also, as $B$ increases, the growth of perturbations occurs over longer times and for smaller wavenumbers $k$, resulting in a much higher amplitude. Of course, it should be expected that the linearized analysis eventually breaks down so we next present, in turn, experimental and numerical results.

{\it Experiments.} -- For the experiments, we use two silicone oils, one with density $\rho$ = 970 kg/m$^{3}$, surface tension $\gamma$ = 18 mN/m, the kinematic viscosity $\nu$ = 1000 cSt, and the other one with density $\rho$ = 950 kg/m$^{3}$, surface tension $\gamma$ = 19 mN/m, the kinematic viscosity $\nu$ = 20 cSt.  We create substrate curvature by mechanically fixing two sides of a flexible polycarbonate sheet, as shown in Fig.~\ref{Fig:setup}b. The curvature is adjusted by precise displacements of two linear stages installed at the edges of the deformable plate. 

To make a uniform thin liquid film, we first coat silicone oil on top of the substrate by using a spin coater. Then, the liquid film layer is maintained on a horizontal table until the film thickness is uniform, which is validated by measurements with a confocal laser scanning microscope with fluorescent dye and an optical interferometry device. The error in the position of the horizontal axis of the aligned table is less than 0.1$^{\circ}$, which is calibrated by a digital protractor (PRO3600).

When a uniform thin liquid layer is obtained, we impose a curvature to the substrate and turn it upside down. By varying the initial film thickness, $h_{i}$, of silicone oil and the curvature ($1/R$) of the substrate, we vary the aspect ratio in the range $2\times 10^{-5}<\delta< 2.2\times 10^{-3}$ and the modified Bond number in the range $4<B<164$. 

To measure the evolution of the film thickness, we use an optical interferometry device (Fiber optic spectrometer, OceanOptics). The device is integrated with a light source (Tungsten halogen lamp, LS-1-LL) and a CCD array-based spectrometer, which measures wavelengths between 200-1100 nm. Due to the refractive index of the film, there is an optical path difference $\Delta$ between the reflected and refracted wavelengths. By Snell's law, $\Delta$ $\cong$ 2$n_{f}h$, we can measure the dimensionless film thickness, $h(\theta,t)$, where $n_{f}$ is a refractive index of the thin film; in our set-up the inclination angle of the  light source is 90$^{\circ}$ ($\theta=0$). 
 Also, the  refractive index of the silicone oil is 1.403 and the optical path is calculated by the optical interference pattern that is obtained from the spectrometer.

We measure the evolution of the film thickness at the center of the geometry ($\theta=0$), as shown in the dashed box of Fig.~\ref{Fig:setup}b. Typical measurements of the film thickness $h(0,t)$ for modest values of $B$ are presented in Fig.~\ref{Fig:experiments}b. Based on the theoretical results, we expect that for $B < 8$, the film thickness near the top should decrease monotonically; this is demonstrated by, for example, the profile of $B = 7.73$ in Fig.~\ref{Fig:experiments}b. Indeed, other experimental results with $B < 8$ confirm similar behaviour: within this stable regime, the gravitational Rayleigh-Taylor instability is always suppresed due to the curvature of the substrate. 

\begin{figure}[htb]
  \centering
  \includegraphics[width=0.9\textwidth]{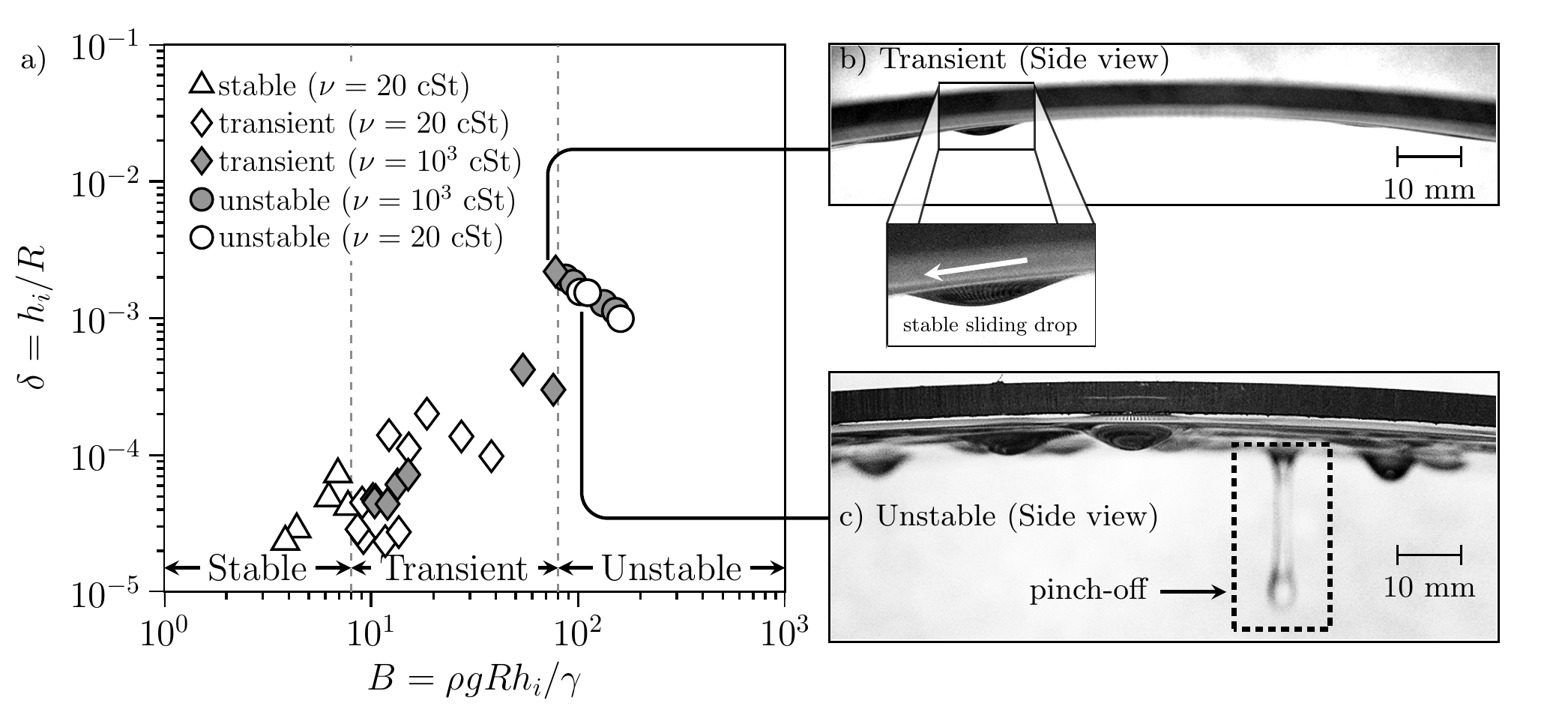}
  \caption{(a) Experimental results in the $(B, \delta)$-plane and classified as stable, transient, or unstable. Two critical values of the Bond number are indicated: $B = 8$ and $B = 80$. Experimental images of the side view in the case of (b) transient and (c) unstable flows.\label{Fig:profiles}}
\end{figure}

For values of $B>8$ and $B = O(1)$, the theory predicts an initial transient growth of perturbations followed by eventual decay. This, too, is confirmed by the profiles of Fig.~\ref{Fig:experiments}b. Within this transient regime, we observe the formation of small droplets on the free surface, and their motion is followed by observing the side and top views using DSLR cameras (Nikon D5100 and D90 with AF-S DX Zoom-Nikkor 18-55mm f/3.5-5.6G lens). As shown in the side view of Fig.~\ref{Fig:profiles}b, the diameter of the sliding droplet is about 12 mm. These droplets slide in the angular direction, but remain attached to the interface (i.e. no dripping). Their size can be estimated by balancing surface tension and gravity, i.e. the typical dimension is $2\pi\sqrt{2}\ell_c$ where $\ell_c=\sqrt{\gamma / \rho g}$ is the capillary length. The sliding sessile droplet size is identical with the wavelength of the fastest growing mode in the classical Rayleigh-Taylor instability problem~\cite{fermigier1992two}. 

If we continue to increase $B$ ($B \gg 1$ in the theory and $B > 80$ in the experiments), the thin film will destabilize. As shown in Fig.~\ref{Fig:profiles}c and the supplementary movie B, this regime of large values of $B$ is characterized by drops pinching off from the free surface. The diameter of all droplets is again about 12 mm. We further note that for $B > 15$, three dimensional effects were experimentally observed in the form of wave patterns in the axial direction ($z$), which can be seen in the supplementary video A.

Thus we see that each experimental result can be classified as stable, transient, or unstable, and this is summarized in the $(B, \delta)$ phase plane shown in Fig.~\ref{Fig:profiles}a. We further note the $B = O(1)$ transition between curvature-stabilized dynamics and the classical Rayleigh-Taylor instability, first noted through the study of Eq.~\eqref{ThinFilmCurvedSubstrate2}, can be alternatively derived by balancing the time scale for drainage of the thin film, $O(\mu R/(\rho g h_i^2))$, with the time scale for growth of the Rayleigh-Taylor instability, which is $O(\mu \gamma/[(\rho g)^2 h_i^3])$ (de Gennes \emph{et al.}\cite{degennes_book}, p.117). 

\begin{figure}[htb] \centering
\includegraphics[width=0.35\textwidth]{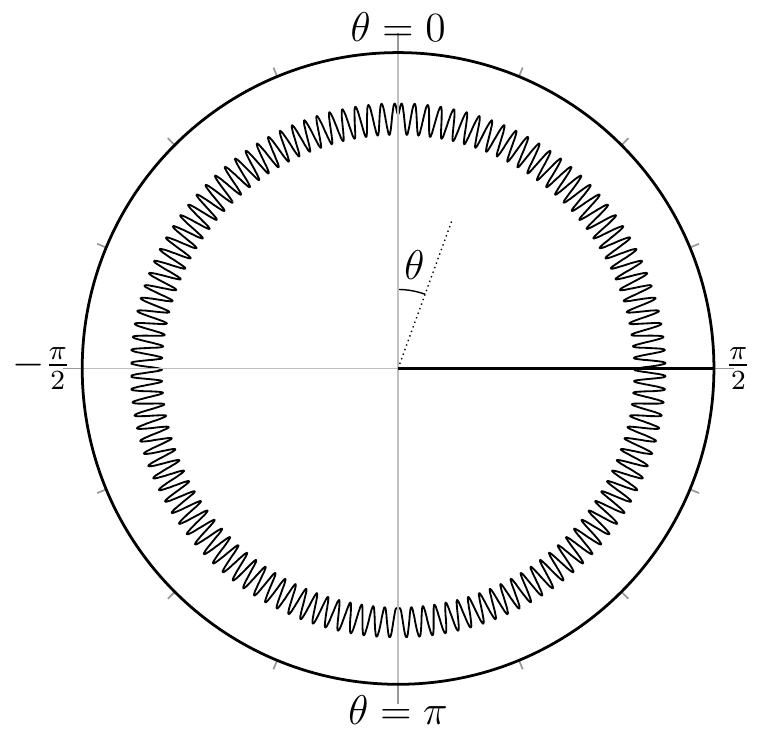} \hspace*{1cm} \includegraphics[width=0.35\textwidth]{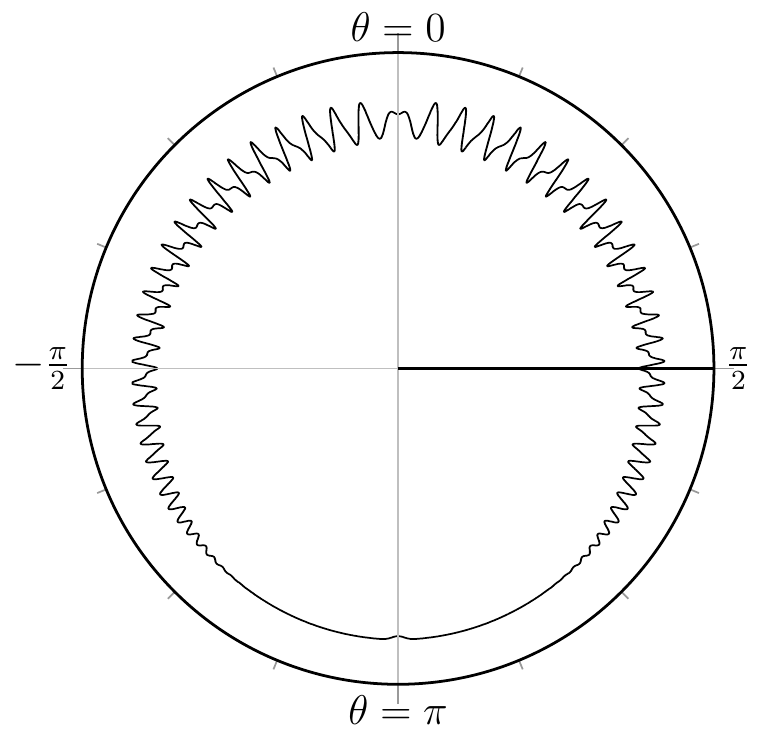} 
\caption{Plots of $H(\theta, t) = \tfrac{1}{D}[h(\theta, t) - h_\text{flat}(\theta, t)]$ from Eq. \eqref{H} at times $t = 0$ (left) and $t = 1$ (right), showing the evolution of an initial sinusoidal disturbance on a thin film in a circular cylinder. The parameters are $B = 20$ and $\delta = 10^{-3}$, with initial condition $h(\theta, 0) = 1 + D\sin(\pi k^* \theta/\sqrt{\delta})$ (same as Fig.~\ref{Fig:experiments}a). The radial scaling of the two profiles has been exaggerated. \label{typical}}
\end{figure}

{\it Numerics.} -- We seek to compare the asymptotic prediction of perturbation growth in Eq. \eqref{etaeq} (only valid near the top of the cylinder) to both the experimental results and the numerical solutions of the full thin film problem in Eq. \eqref{ThinFilmCurvedSubstrate2}. This full problem is solved using finite differences beginning from an initial condition $h(\theta, 0) = 1 + D\sin(\pi k^* \theta/\sqrt{\delta})$. For $B \geq 8$, we choose $k^*$ to be the wavenumber in Eq. \eqref{etaeq} that produces maximal growth; for $B < 8$, we choose $k^* = 2$ (its value at $B = 8$). We then compute the difference
\begin{equation} \label{H}
H(\theta,t) = \frac{1}{D} \left[h(\theta,t) - h_\text{flat}(\theta,t)\right]
\end{equation}
where $h_\text{flat}$ is the solution of \eqref{ThinFilmCurvedSubstrate2} with an initial condition with no perturbation, $D = 0$. Numerical profiles of $h(\theta, t)$ and $h_\text{flat}(\theta, t)$ are shown in Fig.~\ref{Fig:experiments}a, and the corresponding differences, $H(\theta,t)$ are shown in Fig.~\ref{typical}. We see the decay of the initial sinusoidal perturbation (Fig.~\ref{typical}a) as the waves are advected towards the bottom of the cylinder (Fig.~\ref{typical}b).

\begin{figure}[htb]
\includegraphics[width=0.55\textwidth]{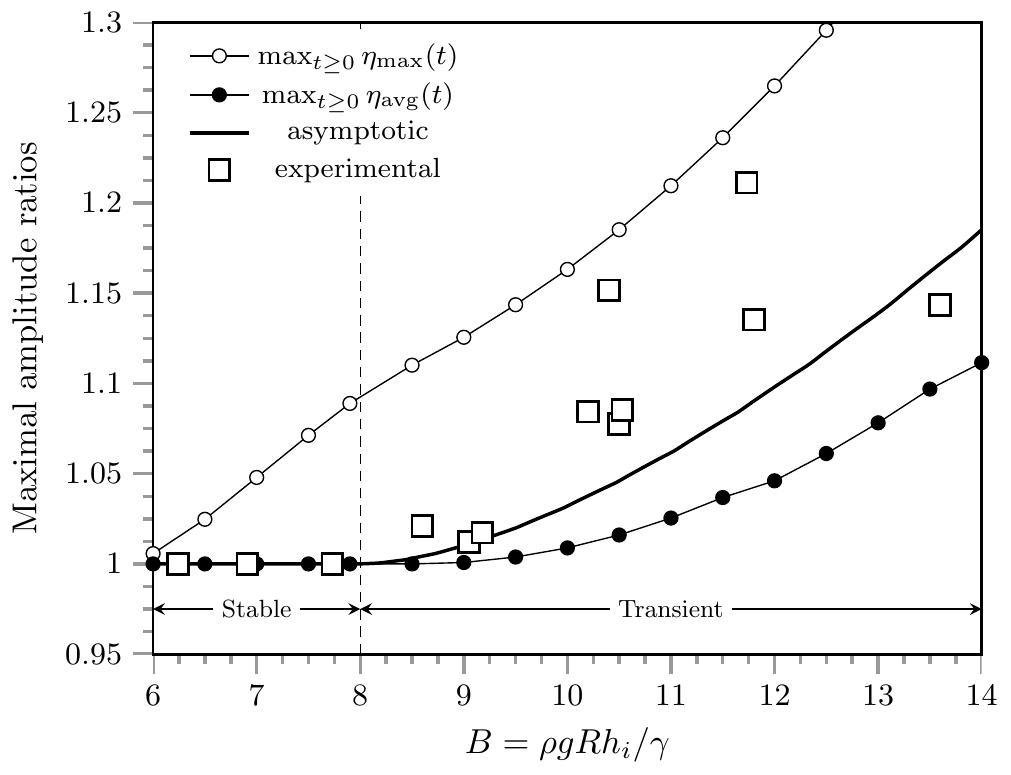}
\caption{Maximal amplitude ratios, $\eta_\text{max}$ (top curve) and $\eta_\text{avg}$ (bottom curve) over all time for the full problem in Eq. \eqref{ThinFilmCurvedSubstrate2} with $\delta = 10^{-3}$, $D = 10^{-2}$, and $k = k^*$. The asymptotically predicted $\max_{t \geq 0} A(k^*, t)$ (thick line) lies between the two numerical results. The experimental results (squares) measure the maximum value of $h(0, t)/h(0, 0)$ (see Fig.~\ref{Fig:experiments}). \label{numerical}}
\end{figure}

To measure the magnitude of the perturbations over the upper half of the cylinder, for each choice of $D$, $k^*$, $B$, and $\delta$, we use two measures: (i) the maximal perturbation, $\eta_\text{max}(t) \equiv \max_{\theta \leq \pi/2} H(\theta, t)$, or (ii) the average amplitude, $\eta_\text{avg}(t)$, of the oscillations in $H$ for $\theta \leq \pi/2$. The maximal values of $\eta_\text{max}(t)$ or $\eta_\text{avg}(t)$ over all time are computed and compared with the maximal value of $A(k^*, t)$ over all time. 
%
The results are shown in Fig.~\ref{numerical}, and demonstrate that the asymptotic prediction of Eq. \eqref{Aeq} lies between the two measures of perturbation instability (for $\delta = 10^{-3}$ and $D = 10^{-2}$). The numerical results agree favourably with the experimental measurements. The critical value of $B = 8$ can be identified clearly in asymptotic, numerical, and experimental results.

{\it Conclusion.} -- 
We have studied the gravitational Rayleigh-Taylor instability for a liquid film under a curved substrate. 
Using theoretical and experimental approaches, we identify two critical conditions to separate different flow regimes, and document how a geometric feature, 
 the substrate curvature, can suppress the Rayleigh-Taylor instability.
Of course, there are other ways to suppress this instability, e.g. different external effects such as temperature gradients~\cite{burgess2001suppression}, applied electric fields~\cite{1986bez}, and mechanical forces~\cite{wolf1970dynamic}.

It is also possible to extend our results to three-dimensional flow situations. For example, we would expect that for three-dimensional substrates where the local curvature is locally parabolic, the Rayleigh-Taylor instability is suppressed. However, nonlocal curvature gradients and other effects of non-axisymmetric topographies (e.g. imposing inlet or outlet conditions for the draining process) may change the flow characteristics. These more general three-dimensional aspects are the subject of ongoing investigations. 

{\it Acknowledgements.} -- We thank R. Goldston, M. Jaworski, R. Kaita, B, Koel, R. Majeski and C. Skinner for helpful conversations about liquid walls as plasma-facing components, and also I. Jacobi for helping in the film thickness measurements. The DOE Fusion Energy Sciences Program is thanked for support of this research via grant DE-SC0008598 to HAS. Finally, we gratefully acknowledge the Oxford-Princeton Collaborative Workshop Initiative for providing an opportunity for this collaborative work.


\providecommand{\noopsort}[1]{}\providecommand{\singleletter}[1]{#1}%

\end{document}